\documentclass[prb,twocolumn,showpacs,preprintnumbers,amsmath,
amssymb,superscriptaddress]{revtex4}

\usepackage{graphicx}
\usepackage{dcolumn}
\usepackage{bm}

\def\Vec#1{\bm{#1}}

\begin{document}


\title{
Comment on arXiv:1105.6233 entitled \\
``Neutron-Inelastic-Scattering Peak by Dissipationless Mechanism in the s$_{++}$-wave State in Iron-based Superconductors'' 
by S. Onari and H. Kontani
}

\author{Yuki Nagai}
\affiliation{CCSE, Japan  Atomic Energy Agency, 5-1-5 Kashiwanoha, Kashiwa, Chiba, 277-8587, Japan}
\affiliation{CREST(JST), 4-1-8 Honcho, Kawaguchi, Saitama, 332-0012, Japan}
\affiliation{TRIP(JST), Chiyoda, Tokyo 102-0075, Japan}
\author{Kazuhiko Kuroki}
\affiliation{Department of Applied Physics and Chemistry, The University of Electro-Communications, Chofu, Tokyo 182-8585, Japan}
\affiliation{TRIP(JST), Chiyoda, Tokyo 102-0075, Japan}


\date{\today}

\maketitle

Recently, Onari and Kontani submitted a paper\cite{OnariNew} which 
criticizes  our recent theoretical study \cite{Nagai} 
on the neutron scattering experiment as a probe for determining the 
superconducting gap in the iron pnictides. In their paper, 
Onari and Kontani have developed a formalism 
in which the imaginary part of the 
dynamical spin susceptibility (Im$\chi_s(\Vec{q},\omega)$) 
in the superconducting state can be more 
accurately calculated especially in the $\omega<2\Delta$ regime, 
where $\Delta$ is the superconducting gap.
In section IIIC of their paper, 
they mention that the conclusions of  our paper 
are ``incorrect based on inaccurate numerical calculation''.
In the present Comment, we show that this in fact is not correct.

\begin{figure}[th]
  \begin{center}
    \begin{tabular}{p{\columnwidth}}
      \resizebox{0.5 \columnwidth}{!}{\includegraphics{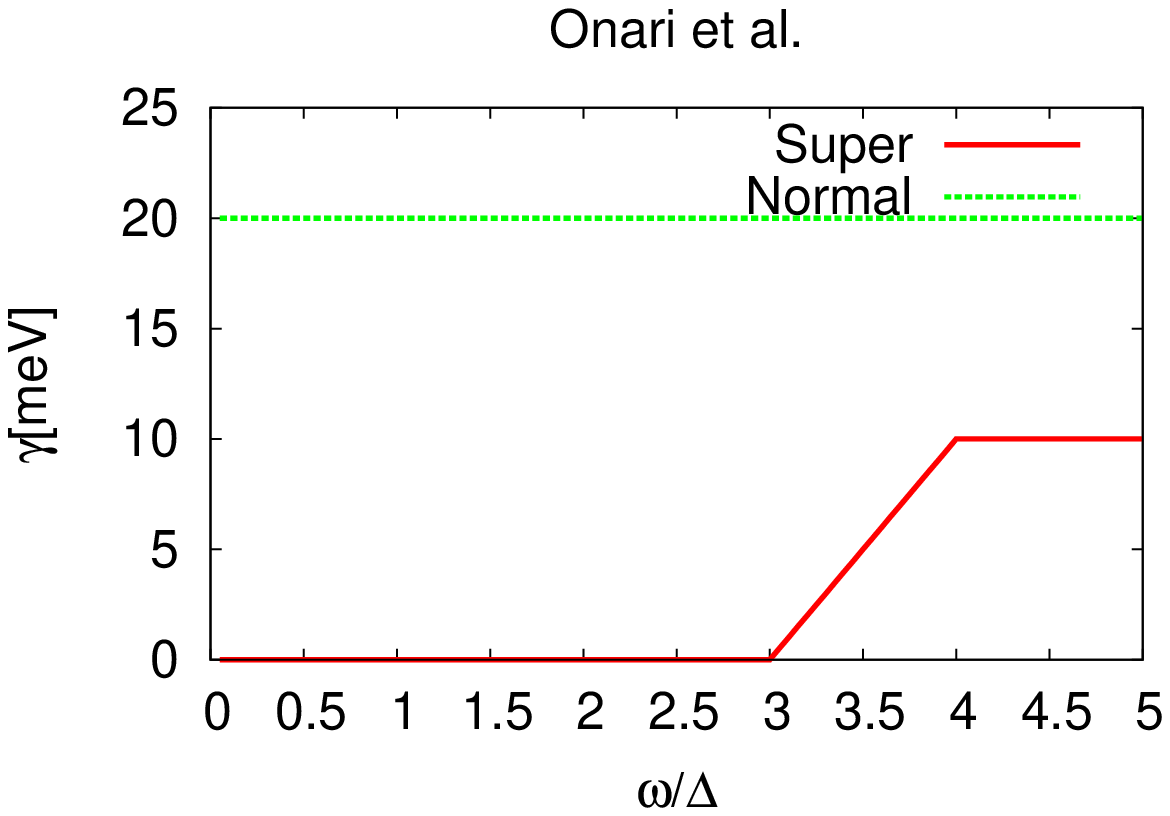}} 
      \resizebox{0.5  \columnwidth}{!}{\includegraphics{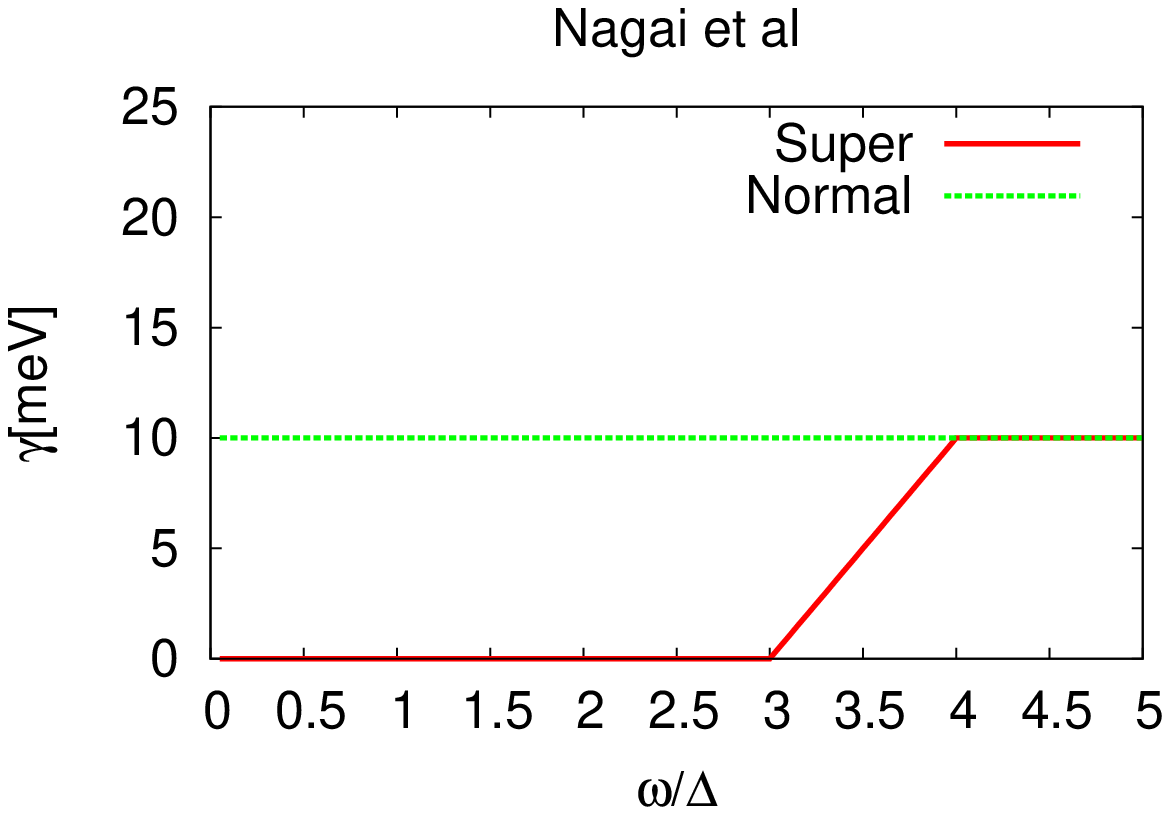}} 
    \end{tabular}
\caption{\label{fig:fig1}
The difference in the quasiparticle damping 
between  Onari {\it et al.} and Nagai {\it et al.}. 
}
  \end{center}
\end{figure}

First of all, Onari and Kontani emphasize the difference of the 
calculation method used, but actually, 
a more important difference between their calculation and ours 
lies in the parameter values taken. The most important 
difference is while we take the strength of the 
quasiparticle damping $\gamma_0$ to be the same 
between the superconducting and the normal states, 
as was done in Onari {\it et al.}'s previous study\cite{OnariOld},
they now take different values for the two states : 
$\gamma_0=20$meV for the normal state and $\gamma=10$meV for the 
superconducting state. The difference of the overall $\gamma$ form 
between the two papers is shown in Fig.~\ref{fig:fig1}.
The larger the difference of $\gamma$ between the 
superconducting and normal states, 
the greater the enhancement (hump in the $s_{++}$ or the resonance 
peak in $s_{\pm}$) of Im$\chi_s$ at $\Vec{q}\simeq (\pi,0)$  
in the superconducting state over that in the normal state, 
in favor of their conclusion that the hump in the $s_{++}$ is large enough 
to explain the experiments while the peak in the $s_{\pm}$ is too large,
as opposed to our conclusion. We do not deny the possibility 
that $\gamma$ values may be different between the two states in the 
actual materials, 
but it is important to notice that the large difference 
among the studies \onlinecite{OnariNew,OnariOld,Nagai} lies here.

On the other hand, Onari and Kontani claim that our conclusion\cite{Nagai} 
that ``the hump in the $s_{++}$ is small for $\Delta=5\sim 10$meV compared to 
the case of $\Delta=25$meV or larger (as was taken in Onari {\it et al}'s 
previous paper\cite{OnariNew})''
is incorrect, based on the observation that Im$\chi_s$ 
in the region $\omega<2\Delta$ is finite, whereas it actually should be 0
in a rigorous calculation. This finiteness is in fact due to the 
approximation adopted in Onari {\it et al.}'s previous paper\cite{OnariOld}.
However, as shown in Fig.4 of their new paper\cite{OnariNew}, 
the result of improving the $\omega<2\Delta$ region 
only results in a {\it suppression of the hump in the $s_{++}$ state}.
Therefore, the improvement in the $\omega<2\Delta$ region does not 
affect our conclusion regarding the enhancement around 
$\omega=2\Delta\sim 3\Delta$ in the $s_{++}$.
As for the $s_{\pm}$ state, 
we do admit that the broadness of the $s_{\pm}$ resonance 
peak in Fig.1(a) of our paper may be due to the approximation, 
but actually we made no statement regarding this broadness, 
as opposed to what is mentioned in section IIIC of Onari-Kontani's paper. 
In fact, we showed another calculation in Fig.1(b) 
 of our paper\cite{Nagai} (which Onari-Kontani hardly mentions about), 
showing a sharper peak. As a common feature between the two cases, 
we mentioned only about the {\it height} of the 
peak in the $s_{\pm}$ state with respect to that in the normal state, 
which is highly dependent on the strength of the quasiparticle 
damping in the normal state.

\begin{figure}[t]
  \begin{center}
      \resizebox{0.8 \columnwidth}{!}{\includegraphics{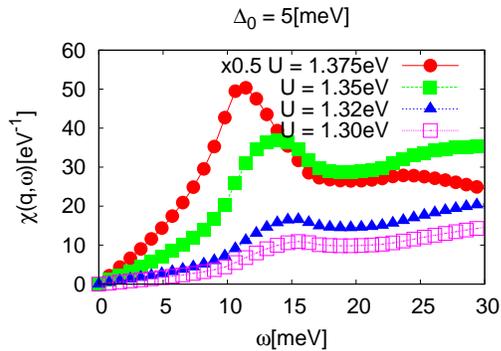}} %
\caption{\label{fig:fig2}
$U$-dependence of Im$\chi_s$ at $\Vec{q}=(\pi,\pi/16)$ 
in the $s_{++}$-wave state with $\Delta = 5$meV and $\gamma_s=10$meV. 
}
  \end{center}
\end{figure}

Also in Fig.5(a) of ref.\onlinecite{OnariNew}, they 
refer to our calculation result (Fig.1(a) of our paper\cite{Nagai})  
and mention that the peak position of the $s_{++}$ hump 
is close to $\omega=2\Delta$, 
whereas it should be closer to $3\Delta$. Here, they have failed to 
mention the value of $U$ in our paper, which is actually $U=1.375$eV.
The peak position is actually dependent on $U$, 
as already seen from the difference between $U=1.3$eV and $U=1.32$eV 
in Fig.5(a) of their paper.
We have performed a calculation for several values of $U$, 
and in fact the peak position systematically varies, as shown 
in Fig.~\ref{fig:fig2}. This variance on $U$ could be to some extent 
due to the approximation adopted here, but in any case, the peak position is 
hardly relevant to our conclusion because in Fig.1(b) of 
our calculation\cite{Nagai}, which Onari-Kontani makes no comment on, 
the peak position is in fact around $\omega=3\Delta$. 
\begin{figure}[t]
  \begin{center}
      \resizebox{0.8 \columnwidth}{!}{\includegraphics{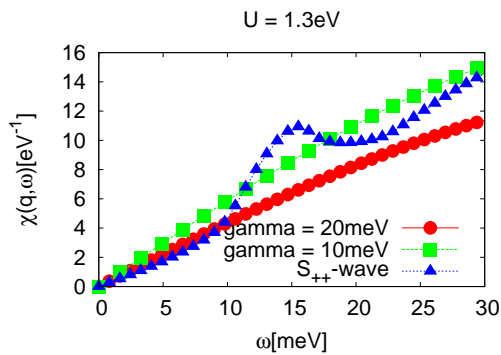}} 
\caption{\label{fig:fig3}
Im$\chi_s$ at $\Vec{q}=(\pi,\pi/16)$ with $U = 1.3$eV.
Blue triangle is for the $s_{++}$ state with $\Delta = 5$meV and 
$\gamma_s=$10meV. 
Red circles and green squares are for the 
normal state with the damping parameter $\gamma_0 = 10$meV and 20meV, 
respectively. 
}
  \end{center}
\end{figure}

To further strengthen this point, 
we once again show in the present Comment, now adopting $U=1.3$eV and 
$\Delta=5$meV, that our conclusion holds as far as we adopt the 
same values of the quasiparticle damping between the normal and 
the superconducting states. In Fig.~\ref{fig:fig3}, we see that 
the enhancement in the $s_{++}$ state above the nomral state is small 
when we adopt $\gamma=10$meV for both states. 
On the other hand, if we adopt $\gamma=20$meV only for the normal 
state, the $s_{++}$ 
hump now appears to be large, consistent with Onari-Kontani's 
claim. However, there does lie a quantitative 
difference between their calculation 
and ours : in the high energy regime of 
$\omega=6\Delta$ or more, Im$\chi_s$ in the superconducting state 
with $\gamma_s=10$meV merges to 
the normal state values with a different $\gamma_0$(=20meV) 
in Onari-Kontani's calculation\cite{OnariNew} 
(Fig.2 and Fig.3 of their paper),  
while in our calculation it merges to the normal state values with the same 
$\gamma_0$(=10meV). In the high energy regime, the effect of 
the superconducting gap should disappear, so that we believe our 
results are more physical. We are not sure about the origin of this 
discrepancy.

To conclude, the main difference between 
Onari-Kontani's calculation\cite{OnariNew} and 
ours (especially Fig.1(b))\cite{Nagai} 
lies in the choice of the quasiparticle damping in the normal state, 
not in the method or the accuracy of the calculation. 
Their claim that the conclusions of  our paper 
are ``incorrect based on inaccurate numerical calculation'' has no basis 
and is in fact not correct.
If we consider the possibility of 
the $\gamma$ values being different between the superconducting and the 
normal states (which we do not deny), then the 
ambiguity of determining the superconducting gap form solely from the 
neutron scattering at $\sim (\pi,0)$ will be greater.
Then our proposal to look into wave vectors other than $(\pi,0)$ 
will have even more important implication\cite{Nagai}.

\end{document}